\documentclass[conference,10pt]{IEEEtran}

\IEEEoverridecommandlockouts

\usepackage{titlesec}
\usepackage{graphicx,kantlipsum,setspace}
\usepackage{amsmath,amssymb,amsthm}
\usepackage{textcomp}
\usepackage{cite}
\usepackage{mathtools}
\usepackage{caption}
\usepackage{subfig}
\usepackage[nolist]{acronym}
\usepackage[hidelinks]{hyperref}
\usepackage{cleveref}
\usepackage{relsize}
\usepackage{float}
\usepackage{epstopdf}
\usepackage{datetime}
\usepackage[ruled,vlined]{algorithm2e}
\usepackage{tabularx}
\usepackage{booktabs}
\usepackage{enumitem}
\usepackage{pgf,tikz}
\usepackage{pgfplots}
\usepackage[strict]{changepage}
\usepackage{ifthen}
\usepackage[fit, breakall]{truncate}
\usepackage{xcolor}
\usepackage[nopostdot,toc,acronym,nomain,nonumberlist]{glossaries}
\usepackage[absolute,overlay]{textpos}

\usetikzlibrary{arrows,positioning} 
\usetikzlibrary{decorations.text}
\usetikzlibrary{arrows.meta}
\usepgfplotslibrary{external}
\tikzsetexternalprefix{external_figs/}
\usepackage{contour}

\graphicspath{{./graphics/}}
\DeclareMathOperator*{\argmin}{arg\,min}
\theoremstyle{definition}

\definecolor{blue33D}{RGB}{0, 0, 175}

\newcommand{\textfonta}{\fontsize{19pt}{21.6pt}\selectfont}
\newcommand{\textfontaB}{\fontsize{20pt}{21.6pt}\selectfont}
\newcommand{\textfontb}{\fontsize{18pt}{16pt}\selectfont}

\definecolor{errorBarC}{RGB}{150,150,150}
\definecolor{c1}{RGB}{215, 196, 255}
\definecolor{c2}{RGB}{89,51,84}
\definecolor{cgrid}{RGB}{215,215,215}
\definecolor{cgrid2}{RGB}{215,215,215}
\definecolor{dark01}{RGB}{75, 75, 75}
\definecolor{dark011}{RGB}{25, 25, 25}
\definecolor{greenC1}{RGB}{13, 128, 82}
\definecolor{greenGF}{RGB}{13, 68, 0}
\definecolor{redC1}{RGB}{209, 6, 60}
\definecolor{redC1c}{RGB}{150, 10, 10}
\definecolor{blueC1}{RGB}{6, 67, 209}
\definecolor{c1}{RGB}{209, 6, 60}
\definecolor{c2}{RGB}{6, 67, 209}
\definecolor{c3}{RGB}{13, 128, 82}
\definecolor{c4}{RGB}{126, 47, 142}
\definecolor{c5}{RGB}{222, 125, 0}
\definecolor{c6}{RGB}{20, 43, 140}
\definecolor{c7}{RGB}{162, 20, 47}

\tikzset{
	gridc/.style= {dotted, cgrid2},
	linew/.style= {line width=1pt, mark options={solid}},
	linew2/.style= {line width=1pt, dashed, dash pattern=on 5pt off 4pt, mark options={solid}},
	linew3/.style= {line width=1pt, dashed, dash pattern=on 1pt off 2pt, mark options={solid}},
	marksz/.style= {mark options={scale=1, fill opacity=1, solid, line width=0.5}},
	marksz2/.style= {mark options={scale=1.4, fill opacity=1, solid, rotate=180, line width=0.5}},
	marksz3/.style= {mark options={scale=1, fill=white, fill opacity=0.5, solid, line width=0.5}},
	linewB/.style= {line width=0.65pt},
	linew2B/.style= {line width=0.75pt},
	markszB/.style= {mark options={scale=1, fill opacity=1, solid, line width=0.5}},
	marksz2B/.style= {mark options={scale=1.4, fill opacity=1, solid, rotate=180, line width=0.5}},
	marksz3B/.style= {mark options={scale=1, fill=white, fill opacity=0.5, solid, line width=0.5}},
	errorlinemarkSty/.style = {mark size=0pt, solid, color=errorBarC},
	errormarkSty/.style = {rotate=90, mark size=0pt, solid, color=errorBarC},
	errorbarSty/.style  = {solid, line width=1pt, opacity=0.9, color=errorBarC, cap=round},
	invisible/.style={opacity=0},
	visible on/.style={alt={#1{}{invisible}}},
	alt/.code args={<#1>#2#3}{%
		\alt<#1>{\pgfkeysalso{#2}}{\pgfkeysalso{#3}} % \pgfkeysalso doesn't change the path
	},
}

\pgfplotsset{
	axisSetup/.style= {axis x line=bottom, axis y line=left, tick align=inside, axis line style={-, line width=1.25pt, color=dark01}},
	short Legend0/.style={%
		legend image code/.code={
			\draw[##1,line width=1pt] plot coordinates {(0pt,0pt) (15pt,0pt)};
		}
	},
	short Legend0B/.style={%
		legend image code/.code={
			\draw[##1,line width=3pt] plot coordinates {(0pt,0pt) (6pt,0pt)};
		}
	},
	short Legend1/.style={%
		legend image code/.code={
			\draw[##1,line width=1pt] plot coordinates {(0pt,0pt) (15pt,0pt)};
		}
	},
	short Legend2/.style={%
		legend image code/.code={
			\draw[##1,linew] plot coordinates {(0pt,0pt) (9pt,0pt)};
			\draw[##1,linew, dashed] plot coordinates {(0pt,2pt) (9pt,2pt)};
			\draw[##1,mark=o, marksz, linew] plot coordinates {(-4pt,1pt)};
		}
	},
	short Legend3/.style={%
		legend image code/.code={
			\draw[##1,linew] plot coordinates {(0pt,0pt) (9pt,0pt)};
			\draw[##1,linew, dashed] plot coordinates {(0pt,2pt) (9pt,2pt)};
			\draw[##1,mark=triangle, linew, marksz2] plot coordinates {(-4pt,1pt)};
		}
	},
	ylabel right/.style={
		after end axis/.append code={
			\node [rotate=90, anchor=north] at (rel axis cs:1,0.5) {#1};
		}   
	}
}

\Crefformat{figure}{#2Fig.~#1#3}
\Crefmultiformat{figure}{Figs.~#2#1#3}{ and~#2#1#3}{, #2#1#3}{ and~#2#1#3}

\newif\ifshowtikz
\showtikztrue
\let\oldtikzpicture\tikzpicture
\let\oldendtikzpicture\endtikzpicture

\renewenvironment{tikzpicture}{%
	\ifshowtikz\expandafter\oldtikzpicture%
	\else\comment%   
	\fi
}{%
	\ifshowtikz\oldendtikzpicture%
	\else\endcomment%
	\fi
}

\makeatletter
\patchcmd{\@makecaption}
{\scshape}
{}
{}
{}
\makeatletter
\patchcmd{\@makecaption}
{\\}
{.\ }
{}
{}
\makeatother
\def\tablename{Table}
\newcommand{\thickbar}[1]{\mathbf{\bar{\text{$#1$}}}}

\begin{document}

\author{Bashar Tahir, Stefan Schwarz, and Markus Rupp \\
	
	\thanks{Bashar Tahir and Stefan Schwarz are with the Christian Doppler Laboratory for Dependable Wireless Connectivity for the Society in Motion. The financial support by the Austrian Federal Ministry for Digital and Economic Affairs and the National Foundation for Research, Technology and Development is gratefully acknowledged.}

	Institute of Telecommunications, Technische Universit\"{a}t Wien, Vienna, Austria 
%	Email: \{bashar.tahir, stefan.schwarz, markus.rupp\}@tuwien.ac.at
}

\title{Impact of Channel Correlation on Subspace-Based Activity Detection in Grant-Free NOMA}

\maketitle
\begin{abstract}
In this paper, we consider the problem of activity detection in grant-free code-domain non-orthogonal multiple access (NOMA). We focus on performing activity detection via subspace methods under a setup where the data and pilot spreading signatures are of different lengths, and consider a realistic frame-structure similar to existing mobile networks. We investigate the impact of channel correlation on the activity detection performance; first, we consider the case where the channel exhibits high correlation in time and frequency and show how it can heavily deteriorate the performance. To tackle that, we propose to apply user-specific masking sequences overlaid on top of the pilot signatures. Second, we consider the other extreme with the channel being highly selective, and show that it can also negatively impact the performance. We investigate possible pilots' reallocation strategies that can help reduce its impact.
\end{abstract}

%\begin{IEEEkeywords}
%	\ac{IRS}, \ac{NOMA}, Gamma moments matching, outage analysis, interference cancellation. 
%\end{IEEEkeywords}

\IEEEpeerreviewmaketitle

\section{Introduction}
Next-generation mobile networks will provide connectivity to a large number of simultaneously accessing devices \cite{Giordani20}, owing specifically to the expansion of \ac{MTC}. As current systems are mainly grant-based, the \acp{UE}' access is managed centrally by \acp{BS}. This can cause large control overhead for setting up the connections and high access-latency, when a large number of \acp{UE} attempt to access the network simultaneously \cite{Samad19a}. In order to address that, grant-free access has been proposed \cite{Liu18a, Wunder14}. It allows the \acp{UE} to transmit their data on their own, without having to be explicitly scheduled by the \ac{BS}. However, since the \acp{UE} transmit on their own, it can happen that multiple \acp{UE} choose to contest the same resources, thus causing a collision of the transmitted packets. At this point, the framework of \ac{NOMA} comes into action, as it provides the capability to manage the multi-user interference. The combination, grant-free \ac{NOMA}, has received wide attention in the literature and has shown the capability to resolve the collisions and support a large number of \acp{UE} accessing the network in a grant-free manner \cite{Wang16a,Shahab20,Dai18a}.

Since the \ac{BS} is not aware of which \acp{UE} are active at a certain time, it has to perform user activity detection first, before it proceeds with channel estimation and data detection. Given the short-packet nature of \ac{MTC} traffic and its sporadicity, the subset of \acp{UE} active at a certain time is typically smaller than the total number of \acp{UE} available. This sparsity has motivated the application of \ac{CS}-based methods, where the orthogonal matching pursuit algorithm and its extensions have been proposed in the literature, such as in \cite{Wang16,Zhang19}. Another category of algorithms are those based on the estimated sample autocorrelation matrix, where subspace methods, such as \ac{MUSIC} can be used \cite{Hasan21,Cheng202}. In \cite{Chao17} joint activity and data detection using approximate message passing and expectation maximization is proposed. Other algorithms such as expectation propagation has been applied in \cite{Jinyoup19}. 
Deep learning was also considered for this problem, as in \cite{Kim20}.

In this work, we consider activity detection in uplink code-domain \ac{NOMA} using subspace methods; namely, the \ac{MUSIC} algorithm. Different from the majority of the mentioned works, we assume that the data and pilots employ different spreading sequences (or signatures). Typically, relatively long sequences are employed for the activity detection and channel estimation in order to support a large number of devices; however, applying these long sequences to the data part can be highly inefficient, as it can substantially reduce the spectral efficiency of the data transmission. Therefore, for the data part we employ short spreading, and the long sequences are only employed for the pilots. Moreover, we apply the activity detection in the context of a  frame-structure that is similar to the \ac{LTE}/\ac{5G} uplink. Our focus is on the impact that the channel correlation has on the performance of activity detection. In the first part, we show that strong time-frequency correlation of the channel can prevent successful detection of the active set of \acp{UE}. In order to address that, we propose overlaying the pilot sequences with user-specific masking sequences, which results effectively in a decorrelation of the channel. Then, in the second part, we consider the influence of strong time-frequency selectivity, and show how it can also negatively impact the performance. To manage the strong selectivity, we investigate repositioning the pilot signatures over the time-frequency grid.

\section{System Model}
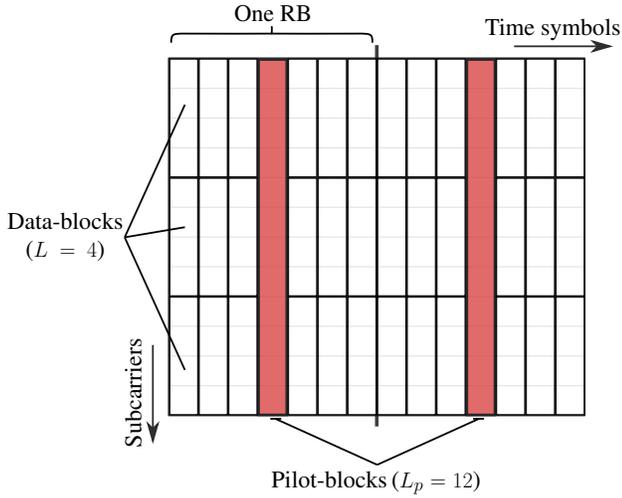
\begin{figure}
	\centering
	\resizebox{1\linewidth}{!}{
		\begin{tikzpicture}
			\node[,] (image) at (0,0) {\includegraphics{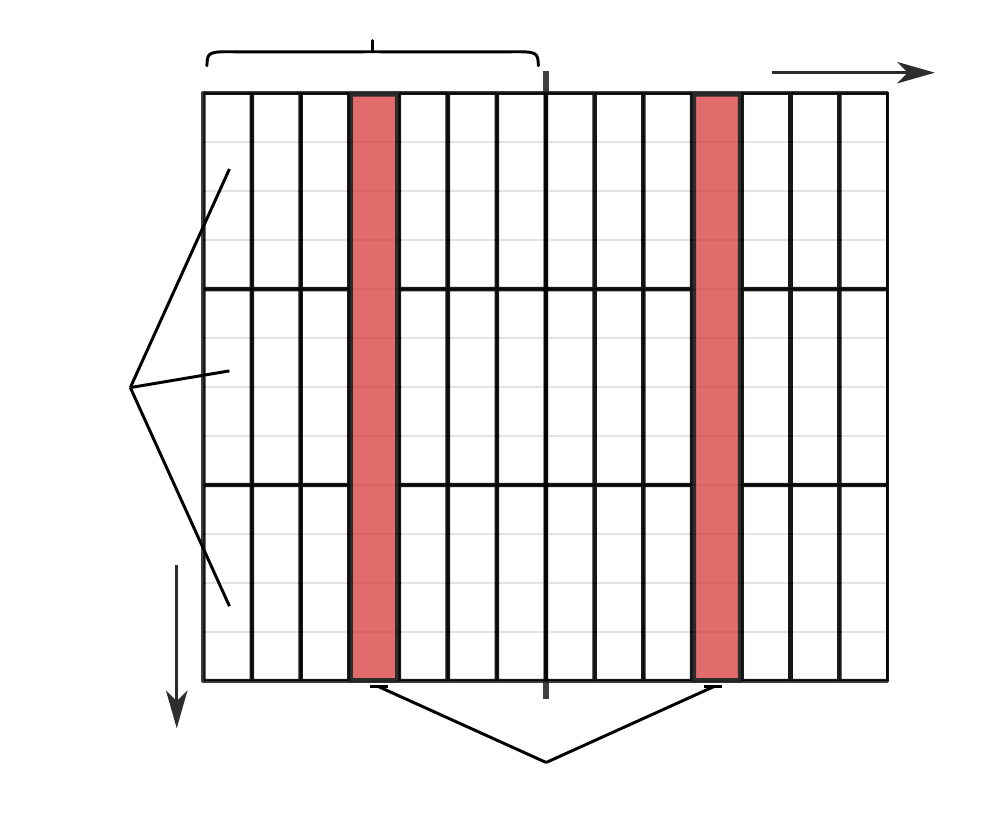}};
			\node at (-7.9, 0.45) [text width=3.5cm, align=center] {\textfonta  Data-blocks ($L = 4$) \par};
			\node at (0.77, -6.4) [text width=6cm, align=center] {\textfonta Pilot-blocks ($L_p = 12$)};
			\node at (5.7, 6.25) {\textfonta Time symbols};
			\node at (-2.1, 6.7) {\textfonta One RB};
			\node[rotate=90] at (-6, -3.9) {\textfonta Subcarriers};
		\end{tikzpicture}
	}
	\caption{Considered frame-structure with 12 subcarriers and 14 OFDM symbols.}
	\vspace{-7mm}
	\label{figc4p1:04}
\end{figure}
We consider a synchronous code-domain \ac{NOMA} uplink consisting of a total of $K$ \acp{UE} transmitting via \ac{OFDM}. We utilize the \ac{LTE}/5G-like frame-structure depicted in \Cref{figc4p1:04}, comprising of two \acp{RB} in time, with each \ac{RB} consisting of 12 subcarriers and 7 \ac{OFDM} symbols. The transmitted frame consists of data-blocks resulting from the \ac{UE} spreading its data-symbols with a spreading signature of length $L$, and also pilots-blocks, where a pilot signature of length $L_p$ is inserted. Let the number of active \acp{UE} at a certain instant be $K_a$ out of $K$, the received baseband signal at the \ac{BS} after \ac{OFDM} demodulation at pilot-block $i$ is given by
\begin{align}\label{eqc4p1:01}
\mathbf{y}_{p, i} = \sum_{k = 1}^{K_a} \sqrt{L_pP_k}\,\mathbf{a}_k h_{k, i} + \mathbf{n}_{p,i},\quad i = 1,2,\dots, B_p,
\end{align}
where $P_k$ and $\mathbf{a}_k \in \mathbb{C}^{L_p}$ are the transmit power and pilot signature of \ac{UE}-$k$, respectively, $h_{k, i} \in \mathbb{C}$ is the channel coefficient of \ac{UE}-$k$ at pilot-block $i$, $\mathbf{n}_{p,i} \in \mathbb{C}^{L_p}$ is the zero-mean Gaussian noise with variance $\sigma_{\mathbf{n}}^2$ at pilot-block $i$, and $B_p$ is the total number of pilot-blocks. Here, for the start, it is assumed that the fading is flat across each pilot-block; hence, it is given as a scalar. Let $\mathbf{A} = [\mathbf{a}_1,\,\mathbf{a}_2,\,\dots,\,\mathbf{a}_{K_a}]$, $\mathbf{h}_i = [\sqrt{L_pP_1}h_{1, i},\,\sqrt{L_pP_2}h_{2, i},\,\dots,\,\sqrt{L_pP_{K_a}}h_{{K_a}, i}]^T$, we can write \eqref{eqc4p1:01}  equivalently as
\begin{align}
\mathbf{y}_{p, i} = \mathbf{A}\mathbf{h}_i + \mathbf{n}_{p,i}.
\end{align}
Under such a system model, the autocorrelation matrix of the received signal over the pilots-blocks (over index $i$) is
\begin{align}\label{eqc4p1:03}
\mathbf{R}_{\mathbf{y}_p} = \mathbb{E}\{\mathbf{y}_{p}^{}\mathbf{y}_{p}^H\} =  \mathbf{A}\mathbf{R}_{\mathbf{h}}\mathbf{A}^H + \sigma_{\mathbf{n}}^2 \mathbf{I},
\end{align}
where $\mathbf{R}_{\mathbf{h}} = \mathbb{E}\{\mathbf{h}\mathbf{h}^H\}$. As can be seen, the autocorrelation matrix consists of two components, corresponding to the signal and noise parts. The idea behind subspace methods is, as long as $0 < K_a <L_p$, then the eigenspace of the autocorrelation matrix can be divided into signal-plus-noise and noise-only subspaces. 
%The active signatures live in the signal subspace, and theoretically, they have zero contribution to the noise subspace.
 Based on this, it is possible to tell which signature is active based on how much energy it has in the noise subspace. To elaborate further, consider the eigenvalue decomposition of $	\mathbf{R}_{\mathbf{y}_p}$ given by
\begin{align}
\mathbf{R}_{\mathbf{y}_p} = \mathbf{U}\mathbf{\Lambda}\mathbf{U}^H,
\end{align}
with $\mathbf{\Lambda} = \text{diag}(\lambda_1, \lambda_2,\dots,\lambda_{L_p})$ being the matrix of eigenvalues with order $\lambda_1 > \lambda_2 > \dots > \lambda_{L_p}$, and $\mathbf{U} = [\mathbf{u}_1, \mathbf{u}_2, \dots, \mathbf{u}_{L_p}]$ is the matrix of the corresponding eigenvectors. Given that $K_a$ \acp{UE} are active, the noise subspace $\mathbf{U}_{\mathbf{n}}$ is given by the collection of eigenvectors corresponding to the smallest $L_p - K_a$ eigenvalues, i.e.,
\begin{align}\label{eqc4p1:05}
\mathbf{U}_{\mathbf{n}} = [\mathbf{u}_{K_a+1},\mathbf{u}_{K_a+2},\dots, \mathbf{u}_{L_p}].
\end{align}
The \ac{MUSIC} spectrum is then calculated as
\begin{align}\label{eqc4p1:06}
M(k) = \frac{1}{\lVert \mathbf{U}_{\mathbf{n}}^H \mathbf{a}_k \rVert^2}, \quad k = 1,2,\dots,K.
\end{align}
The $K_a$ signatures with highest $M(k)$ are then declared active. 
%All the other signatures that are not active occupy a portion of the noise-subspace and therefore will result in small $M(k)$.
In practice, we do not have access to the true autocorrelation matrix, but rather we estimate it from the received pilot-blocks. The sample autocorrelation matrix is calculated as
\begin{align}\label{eqc4p1:07}
\hat{\mathbf{R}}_{\mathbf{y}_p} = \frac{1}{B_p} \sum_{i=1}^{B_p} \mathbf{y}_{p, i}\mathbf{y}_{p, i}^H.
\end{align}
If multiple receive antennas are available at the \ac{BS}, then this estimate can be further improved by averaging over the pilot-blocks from all of the receive antennas.
\vspace{-2mm}
\subsection{Estimation of $K_a$}
\vspace{-1mm}
In order to determine the noise subspace in \eqref{eqc4p1:05} and pick the strongest $K_a$ signatures from \eqref{eqc4p1:06}, we need to know $K_a$ a priori; however, that is not possible in grant-free access, since the \ac{BS} is not aware of how many \acp{UE} are active at a certain time. A similar issue also exists in other activity detection methods, such as in \ac{CS}, where the sparsity level of the problem needs to be known. Therefore, $K_a$ has to be estimated from the received signal as well, and here, we utilize the \ac{BIC} \cite{Schwarz78,Stoica04}.
%Other information criteria also exist, such as the \ac{AIC}\cite{Akaike74}; but \ac{BIC} generally is preferred, as it is a consistent selector \cite{Dziak19}, i.e., as the number of available samples (pilot-blocks) increases, the probability of choosing the correct model (with correct $K_a$) approaches $100\%$. However, as argued in \cite{Dziak19}, this inconsistency might not necessary be a flaw in \ac{AIC}, as it can provide an advantage for certain problems compared to \ac{BIC}. In our own testing in the context of activity detection in grant-free \ac{NOMA}, we found that \ac{BIC} performs better, and therefore this is what we adopt next.
Note that \ac{BIC} has been applied for \ac{NOMA} activity detection in \cite{Hasan21}. Compared to that work, we formulate it here for the case where the noise power is known at the \ac{BS}, in a fashion similar to \cite{Ermolaev2001}. 
%The \ac{BIC} is defined as
%\begin{align}\label{eqc4p1:08}
%\textrm{BIC}(K_a) \triangleq -\log f\big(\mathbf{y}_p; \hat{\boldsymbol{\theta}}(K_a)\big) + \frac{1}{2}|\hat{\boldsymbol{\theta}}(K_a)|\ln B_p,\quad K_a = 1, 2, \dots, L_p-1,
%\end{align}
%where $f\big(\mathbf{y}_p; \hat{\boldsymbol{\theta}}(K_a)\big)$ is the likelihood function of the received pilot-blocks, evaluated at the \ac{ML} estimate of the underlying distribution parameters $\hat{\boldsymbol{\theta}}(K_a)$ under the assumption of $K_a$ active components, and $|\hat{\boldsymbol{\theta}}(K_a)|$ is the number of parameters. 
Under the assumption of zero-mean received signal and the samples being i.i.d, the \ac{BIC} for $K_a$ active \acp{UE} is given by \cite{Ermolaev2001}
\begin{align}\label{eqc4p1:09}
\begin{split}
\textrm{BIC}(K_a) = B_p \sum_{k = 1}^{L_p}\bigg(\frac{\hat{\lambda}_k}{\thickbar{\lambda}_k} +  \ln\thickbar{\lambda}_k \bigg) + B_pL_p\ln\pi \\ +\frac{K_a}{2}\Big(2L_p - K_a\Big)\ln B_p,
\end{split}
\end{align}
where $\hat{\lambda}_k$ is the $k$-th eigenvalue of the estimated sample autocorrelation matrix $\hat{\mathbf{R}}_{\mathbf{y}_p}$, and we set $\thickbar{\lambda}_k$ as
\begin{align}
\thickbar{\lambda}_k = 
\begin{cases}
\hat{\lambda}_k, \quad k = 1,2,\dots,K_a, \\
\sigma_{\mathbf{n}}^2, \quad k = K_a{+}1,...,L_p.
\end{cases}
\end{align}
That is, for the current $K_a$ under investigation, the eigenvalues up to $K_a$ are set equal to their \ac{ML} estimate which we simply obtain from an eigendecomposition of the sample autocorrelation matrix, while the remaining eigenvalues are supposed to belong to the noise-only subspace and thus are set equal to the known noise power. Our estimate of $K_a$ is then the one that minimizes the \ac{BIC}, i.e.,
\begin{align}
\hat{K}_a = \argmin_{K_a} \text{BIC}(K_a).
\end{align}
%It can happen that at a certain time, no \ac{UE} is active (i.e., $K_a =0$), which we would not be able to tell from \eqref{eqc4p1:08}. Typically, \acp{BS} have a certain threshold for decodability; signals with powers below that threshold are considered too weak to be detected anyway. Therefore, one way to tell whether there is an activity at all, is to compare the strongest estimated eigenvalue to that threshold. Let the threshold be $\sigma^2_{\text{min}}$, then $K_a$ is estimated as
%\begin{align}
%\hat{K}_a = 
%\begin{cases}
%0, &\quad  \hat{\lambda}_1 < \sigma^2_{\text{min}}, \\
%\argmin_{K_a} \text{BIC}(K_a), &\quad \text{otherwise}.
%\end{cases}
%\end{align}

\subsection{Channel Estimation and Data Detection}
After finding the active set of \acp{UE}, channel estimation and data detection is performed. Let the matrix of estimated active signatures be $\hat{\mathbf{A}}$, the \ac{LS} channel estimate at pilot-block $i$ is given by
\begin{align}
	\hat{\mathbf{h}}_i =  (\hat{\mathbf{A}}^H\hat{\mathbf{A}})^{-1}\hat{\mathbf{A}}^H\mathbf{y}_{p, i}.
\end{align}
Once the channel estimates are obtained over all pilot-blocks, linear interpolation (and also extrapolation) is performed to obtain the channel for the whole time-frequency grid. 

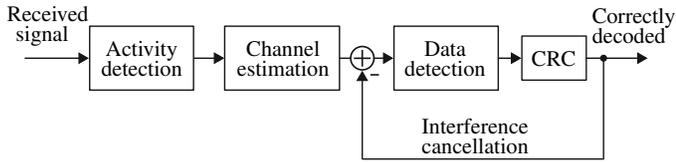
\begin{figure}[t]
	\centering
	\resizebox{1\linewidth}{!}{
		\begin{tikzpicture}
			\def\xshif{0.6}
			\node[,] (image) at (-\xshif,0) {\includegraphics{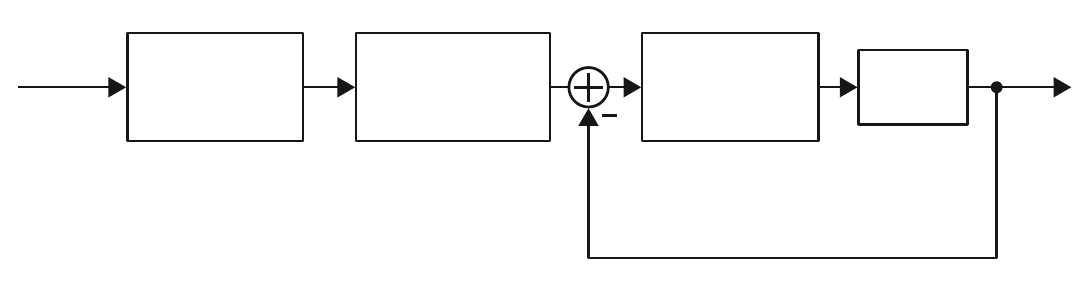}};
			\node at (-9, 1.9) [text width=2cm, align=center]{\textfontb Received\\signal\par};
%			\node at (-5.15, -0.5) [text width=6cm, align=center]{\textfontb Only in grant-free access};
			\node at (-5.55-\xshif, 0.96) [text width=3cm, align=center]{\textfontb Activity detection\par};
			\node at (-1.5-\xshif, 0.96) [text width=3cm, align=center]{\textfontb Channel estimation\par};
			\node at (3.14-\xshif, 0.96) [text width=2.5cm, align=center]{\textfontb Data detection\par};
			\node at (6.25-\xshif, 0.96) [text width=2.5cm, align=center]{\textfontb CRC\par};
			\node at (7, 1.9) [text width=0.5cm, align=center]{\textfontb Correctly\\decoded\par};
			\node at (4-\xshif, -1.25) [text width=3.5cm, align=center]{\textfontb Interference cancellation\par};
		\end{tikzpicture}
	}
	\caption{Receive chain at the \ac{BS}.}
	\label{figc1p1:03}
	\vspace{-2mm}
\end{figure}

The \ac{BS} then proceeds with the data-detection, and in this work, we employ \ac{MMSE} equalization with \ac{PIC} \cite{Tahir20a}. Here, the IC is parallel \ac{CRC}-based, meaning that all \acp{UE} that are decoded and pass their \ac{CRC} get removed from the received signal. The data detection is then repeated on the cleared-up signal until a maximum number of iterations is reached, or no more \ac{UE} passes the \ac{CRC}. The overall receive chain is illustrated in \Cref{figc1p1:03}.

\section{Impact of Strong Time-Frequency Correlation}
The subspace activity detector was formulated under the assumption that the autocorrelation matrix can be divided into distinguished signal-plus-noise and noise-only subspaces. However, in order for this to work, the autocorrelation matrix of the channel coefficients, i.e., $\mathbf{R}_{\mathbf{h}}$ has to be full rank in order for the whole product of $\mathbf{A}\mathbf{R}_{\mathbf{h}}\mathbf{A}^H$ in \eqref{eqc4p1:03} to have a rank equal to the number of active \acp{UE}. To get a full rank $\mathbf{R}_{\mathbf{h}}$, the channel coefficients across the different pilot-blocks need to be i.i.d., which corresponds to the case where the channel is highly selective from one pilot-block to another; however, typically, the channel is correlated in time and frequency, and therefore neighbouring pilot-blocks, especially in time, are likely to experience a similar channel. In other words, in many situations, the signal received at the pilot-blocks can be highly correlated, and this can be an issue for our detector. To further elaborate on this, let us consider the worst-case scenario where all the \acp{UE} undergo flat-fading in both time and frequency. Under such an assumption, we will have $\mathbf{h}_1 = \mathbf{h}_2 = \dots = \mathbf{h}_{B_p} = \mathbf{h}$, i.e., a constant, resulting in $\mathbf{R}_{\mathbf{h}} = \mathbb{E}\{\mathbf{h}\mathbf{h}^H\} = \mathbf{h}\mathbf{h}^H$. The term $\mathbf{h}\mathbf{h}^H$ is simply an outer product of a vector with itself, i.e., a rank-1 matrix. Our resultant autocorrelation matrix is then given by
\begin{align}
\mathbf{R}_{\mathbf{y}_p} =  \mathbf{A}\mathbf{h}\mathbf{h}^H\mathbf{A}^H + \sigma_{\mathbf{n}}^2 \mathbf{I}.
\end{align}
Consequently, the signal part is rank-1; hence, we can only detect one active \ac{UE}. This can also be observed by looking at the sample autocorrelation matrix in \eqref{eqc4p1:07}. If the samples (pilot-blocks) used are correlated, or in a worst-case scenario identical, then it would be a sum of the same outer product, resulting in a rank-1 estimate of the signal part of the autocorrelation matrix. Note that the  \ac{BIC} expression in \eqref{eqc4p1:09} is derived under an i.i.d. factorization of the likelihood function. Therefore, the correlation between the pilot-blocks also impacts this simplified \ac{BIC} calculation.

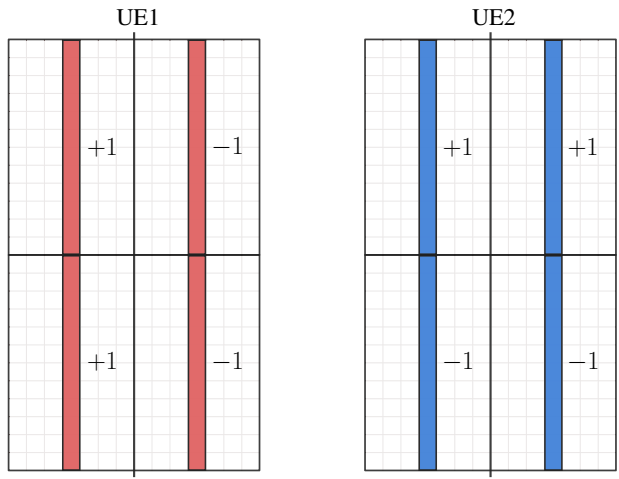
\begin{figure}[t]
	\centering
	\resizebox{1\linewidth}{!}{
		\begin{tikzpicture}
		\def\gloXshif{0.1}
		\node[,] (image) at (0,0) {\includegraphics[width=2.05\linewidth]{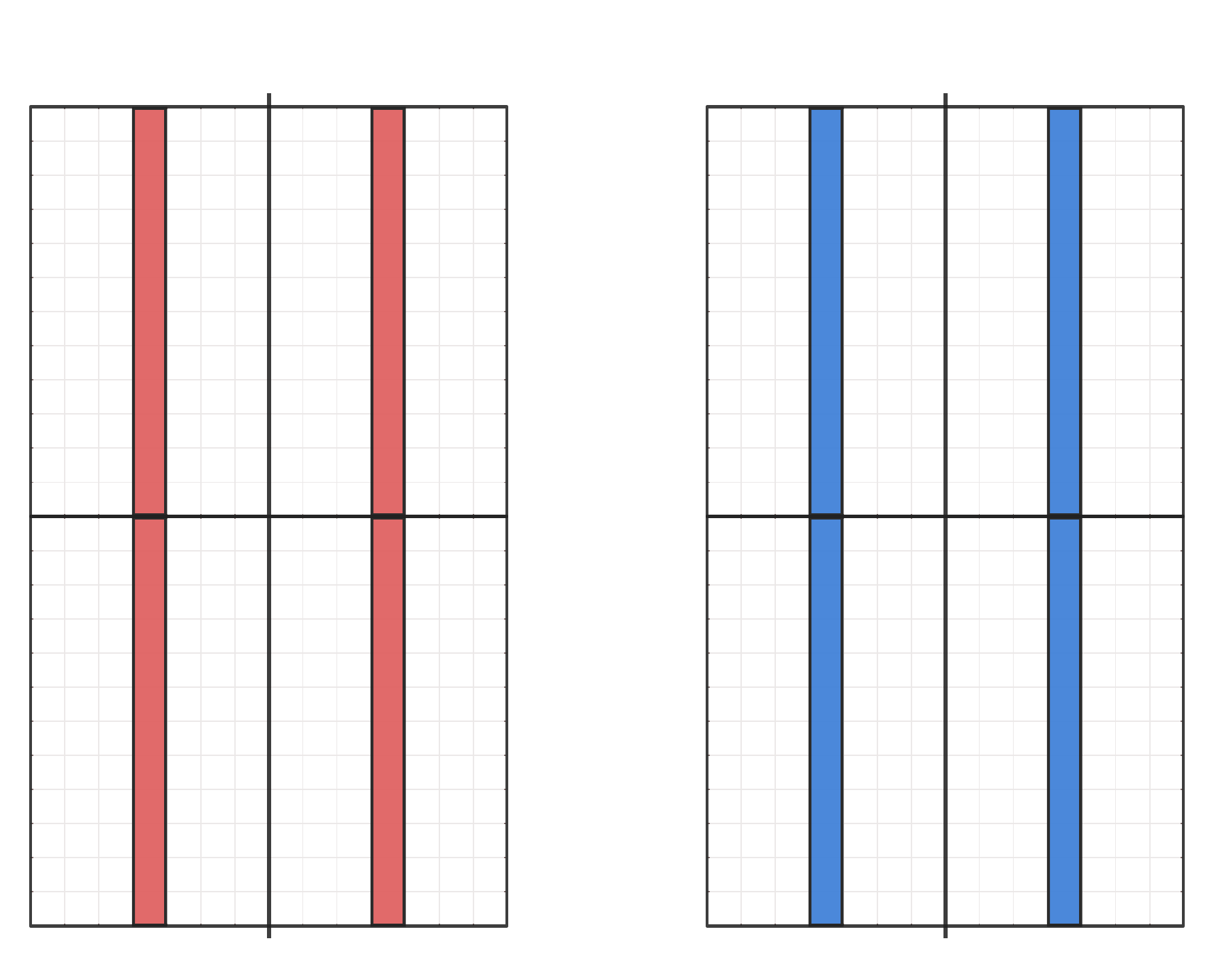}};
		\node at (-5, 6.1) [align=center] {\textfonta  UE1};
		\node at (\gloXshif+-6.1, 2.46) [align=center] {\textfontaB  $+1$};
		\node at (\gloXshif+-6.1, -3.55) [align=center] {\textfontaB  $+1$};
		\node at (\gloXshif+-2.6, 2.46) [align=center] {\textfontaB  $-1$};
		\node at (\gloXshif+-2.6, -3.55) [align=center] {\textfontaB  $-1$};
		\def\xshif{9.95}
		\node at (-5+\xshif, 6.1) [align=center] {\textfonta  UE2};
		\node at (\gloXshif+-6.1+\xshif, 2.46) [align=center] {\textfontaB  $+1$};
		\node at (\gloXshif+-6.1+\xshif, -3.55) [align=center] {\textfontaB  $-1$};
		\node at (\gloXshif+-2.6+\xshif, 2.46) [align=center] {\textfontaB  $+1$};
		\node at (\gloXshif+-2.6+\xshif, -3.55) [align=center] {\textfontaB  $-1$};
		\end{tikzpicture}
	}
	\caption{Masking sequences for two \acp{UE} over 4 \acp{RB}.}
	\vspace{-4mm}
	\label{figc4p1:05}
\end{figure}

Here, to tackle this issue, we propose the use of masking sequences, applied on top of the pilot-blocks. \Cref{figc4p1:05} illustrates the application of masking sequences for \acp{UE} transmitting over four \acp{RB}, where \ac{UE}1 transmits with the red pilot signature, while \ac{UE}2 transmits with the blue one. Instead of transmitting the pilot signatures directly, they are overlaid with a masking sequence. In the considered example, \ac{UE}1 uses the masking sequences $[+1,+1,-1,-1]$, while \ac{UE}2 uses $[+1,-1,+1,-1]$.
The system model now becomes
\begin{align}\label{eqc4p1:14}
\mathbf{y}_{p, i} = \sum_{k = 1}^{K_a} \sqrt{L_pP_k}\,\mathbf{a}_k h_{k, i} m_{k, i} + \mathbf{n}_{p,i},
\end{align}
where $m_{k, i}$ is the $i$-th coefficient of \ac{UE}-$k$ masking sequence applied at the $i$-th pilot-block. Let $\mathbf{m}_i$ be the collection of the masking coefficients across the different \acp{UE} at block $i$, then
\begin{align}
\mathbf{y}_{p, i} = \mathbf{A}\big(\mathbf{h}_i \circ \mathbf{m}_i \big) + \mathbf{n}_{p,i},
\end{align}
where $\circ$ denotes the element-wise Hadamard product. Treating the $\mathbf{m}_i$ as a random process across the pilot-blocks, it can be easily shown that the autocorrelation matrix then is given by
\begin{align}
\mathbf{R}_{\mathbf{y}_p} =  \mathbf{A}\big(\mathbf{R}_{\mathbf{h}} \circ \mathbf{R}_{\mathbf{m}} \big)\mathbf{A}^H + \sigma_{\mathbf{n}}^2 \mathbf{I},
\end{align}
where $\mathbf{R}_{\mathbf{m}} = \mathbb{E}\{\mathbf{m}\mathbf{m}^H\}$. Therefore, by designing the masking sequences to be like an i.i.d. process across the \acp{UE}, then it would be possible to recover a correct estimation of the signal part of the autocorrelation matrix. For example, even if we have $\mathbf{R}_{\mathbf{h}} = \mathbf{h}\mathbf{h}^H$, the product $\mathbf{h}\mathbf{h}^H\circ \mathbf{R}_{\mathbf{m}}$ can still be full rank if $\mathbf{R}_{\mathbf{m}}$ is full-rank, which is achieved by designing the masking sequences random-like. One way to design these sequences is to pick them randomly from the alphabet $\{-1,+1\}$, as illustrated in the example of \Cref{figc4p1:05}. Other designs are also possible, which might enjoy more structure, such as binary Golay sequences \cite{Yu21}. Those masking sequences would be defined in the standard in a similar fashion as the pilot signatures, preferably having a one-to-one correspondence with them. That is, for each pilot signature, there is a corresponding masking sequence.

In the next stage where channel estimation is performed, the effect of employing the masking sequence is removed from the channel estimate. The masking-free estimate is given by
\begin{align}
\hat{\mathbf{h}}_i = \big( (\hat{\mathbf{A}}^H\hat{\mathbf{A}})^{-1}\hat{\mathbf{A}}^H\mathbf{y}_{p, i} \big) \oslash \mathbf{m}_i,
\end{align}
where $\oslash$ denotes the element-wise division. 

\subsection{Example Scenario with Eight Active UEs}
We now evaluate the performance of the activity detection framework with masking sequences. We assume a grant-free system in which the \acp{UE} get preconfigured (e.g., with pilot signatures) by the \ac{BS} during an initial random-access step. Then, for all future transmissions, the \acp{UE} transmit on their own when they have data using the preassigned configuration. This prevents collision of the pilots between different \acp{UE}.

We consider a scenario where there are $K = 32$ possibly active \acp{UE}; however, at a certain time, only $K_a = 8$ \acp{UE} are active. They contest a resources' region of $72$ subcarriers $\times$ $14$ time symbols corresponding to $12$ \acp{RB}, and the \ac{BS} employs $2$ receive antennas. This brings the number of pilot-blocks that are used to calculate the sample autocorrelation matrix to $B_p = 24$ ($12$ for each antenna). The task of the receiver is then find the active pilot set, perform channel estimation, and finally equalize the data part and decode it. We consider a scenario with high time-frequency correlation, by assuming a Pedestrian-A channel model, which has a low \ac{RMS} delay spread of $45$\,ns, and assume the \acp{UE} are static and therefore the channel is time-invariant. We assume that there is a pathloss spread of $\pm 5$\,dB between the \acp{UE}, meaning that the strongest and weakest \acp{UE} can have a gap in the receive power of up to $10$\,dB. Both the data and pilot signatures are from Grassmannian codebooks \cite{Tahir19b}. As for the masking sequences, we construct them randomly once from $\{+1,-1\}$, and then fix them for all the simulation repetitions. The simulation parameters are summarized in \Cref{Tablec4p1:simpara01}. The various processing operations, such as channel coding, modulation, and channel generation are carried out using the Vienna 5G Link-Level Simulator\footnote{Available: \url{https://www.nt.tuwien.ac.at/research/mobile-communications/vccs/vienna-5g-simulators/}.} \cite{Vienna5GLLS}.
\begin{table}[t]
	\begin{center}
		\small
		\def\arraystretch{1.1}
		\begin{tabularx}{1\linewidth}{l |c}
			\hline
			\textbf{Parameter} & \textbf{Value} \\ \specialrule{1pt}{0pt}{1pt}
			Active UEs & $K_a = 8$ out of total $K = 32$ \\
			Contention region & 72 subcarriers $\times$ 14 time symbols\\
			Data signatures & $L = 4$ ($4 \times 16$ Grassmannian codebook) \\
			Pilots signatures & $L_p = 12$ ($12 \times 32$ Grassmannian codebook)\\
			Masking sequences & Randomly generated from $\{+1,-1\}$ \\ 
			Receive antennas & $N_R = 2$ \\
			Center frequency & $2$ GHz \\
			Subcarrier spacing & $15$ kHz \\ Pathloss spread &  $\pm 5$\,dB \\
			Channel model  & Rayleigh, Ped-A ($45$\,ns RMS), velocity = 0 \\
			Modulation & 4-QAM \\
			Channel coding  & Turbo, code-rate $2/3$ \\
			Activity detection  & \ac{BIC} + \ac{MUSIC} \\
			Channel estimation  & \ac{LS} with linear inter/extrapolation \\
			Data detection  & \ac{MMSE}-\ac{PIC} (max. 6 iters) \\
			\hline
		\end{tabularx}
	\end{center}
	\caption{Simulation parameters for the correlated channels.}
	\label{Tablec4p1:simpara01}
\end{table}

\begin{figure}[t]
	\centering
	\includegraphics[width=1\linewidth]{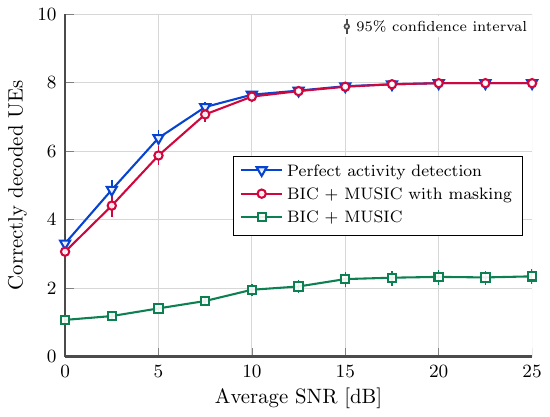}
	\vspace{-6mm}
	\caption{Activity detection with masking sequences.}
	\label{figc4p1:AD01}
	\vspace{-3mm}
\end{figure}

\Cref{figc4p1:AD01} shows the average number of correctly decoded \acp{UE} using the proposed method versus their average \ac{SNR} at the \ac{BS}. The goal here is to have all of the eight active \acp{UE} identified and decoded correctly. The perfect activity detection denotes the case where the \ac{BS} knows exactly which \acp{UE} are active, and therefore it only has to perform channel estimation and data detection. This serves as a baseline for our results. As can be seen in the figure, combining the subspace activity detector with the masking sequences results in a performance that is very close to the case with perfect activity detection. When no masking sequences are employed, then due to the correlation of the channel, it is difficult for the signal part to be constructed properly, thus greatly deteriorating the performance.

Note that if the data part was also spread with the same signatures as the pilots, as commonly done in the literature, then in that case, the received signal from the data part can also be utilized in the construction of the sample autocorrelation matrix. As the data symbols can be assumed i.i.d., they would serve a similar function as the masking sequences, allowing the signal part of the autocorrelation matrix to reach the required rank. However, as we mentioned, spreading the data symbols with long sequences can be spectrally inefficient, and therefore we utilize only the pilots for this task, necessitating the application of the masking sequences.

\section{Impact of Strong Selectivity}
\begin{figure}
	\captionsetup{textfont={normalsize}, labelfont={normalsize}, farskip=15pt, captionskip=2pt}
	\centering
	\subfloat[$0$\,ns \ac{RMS} delay spread.]{
		\centering
		\includegraphics[width=0.8\linewidth]{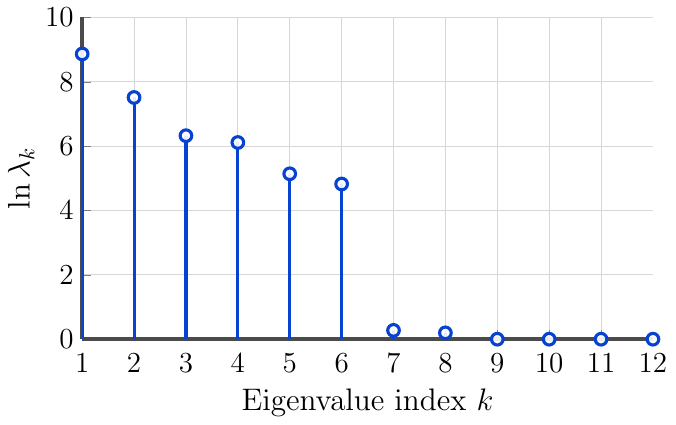}
		\label{figc4p1:eig01}
	}\
	\subfloat[$300$\,ns \ac{RMS} delay spread.]{
		\centering
		\includegraphics[width=0.8\linewidth]{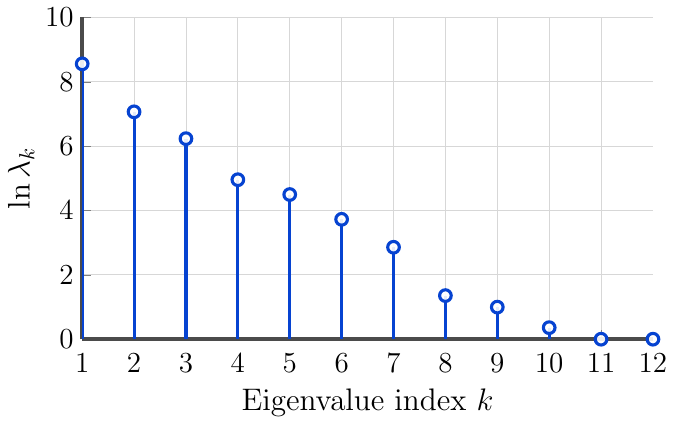}
		\label{figc4p1:eig02}
	}\
	\caption{Eigenvalues of one realization of $\hat{\mathbf{R}}_{\mathbf{y}_p}$  over a \ac{TDL-C} channel for $K_a = 6$ and $L_p =12$.}\label{figc4p1:eig}
%	\vspace{-2mm}
\end{figure}

In the previous section, we discussed the benefit of having a selective channel, as it can reduce the correlation between the received signal at the pilot-blocks, which in turn can improve the detection performance. However, since our transmission is sequence-based, we require the channel to be flat within the spreading interval. Therefore, if the channel is too selective, then our system model would not hold anymore, since different parts of the sequence would experience different channel conditions. This modification of the pilot signature changes from one pilot-block to another, and also it is different for the different \acp{UE}. Under such conditions, the eigenstructure of the autocorrelation matrix no longer reflects the activity of the pilot signatures, but rather a modified version of them.  The more selective the channel is, the more corrupt the estimated autocorrelation matrix will be. In \Cref{figc4p1:eig01}, the eigenvalues of a realization of the autocorrelation matrix for a \ac{TDL-C} channel with no selectivity is shown for $K_a = 6$ and $L_p = 12$. A clear distinction can be seen between the signal part and the noise part in terms of the magnitude of the eigenvalues. On the other hand, \Cref{figc4p1:eig02} shows the case with $300$\,ns \ac{RMS} delay spread. As can be seen, the subspaces are not very clearly separable, which makes it challenging for our subspace detector.

\begin{figure}[t]
	\centering
	\resizebox{0.9\linewidth}{!}{
		\begin{tikzpicture}
		\node[,] (image) at (0,0) {\includegraphics{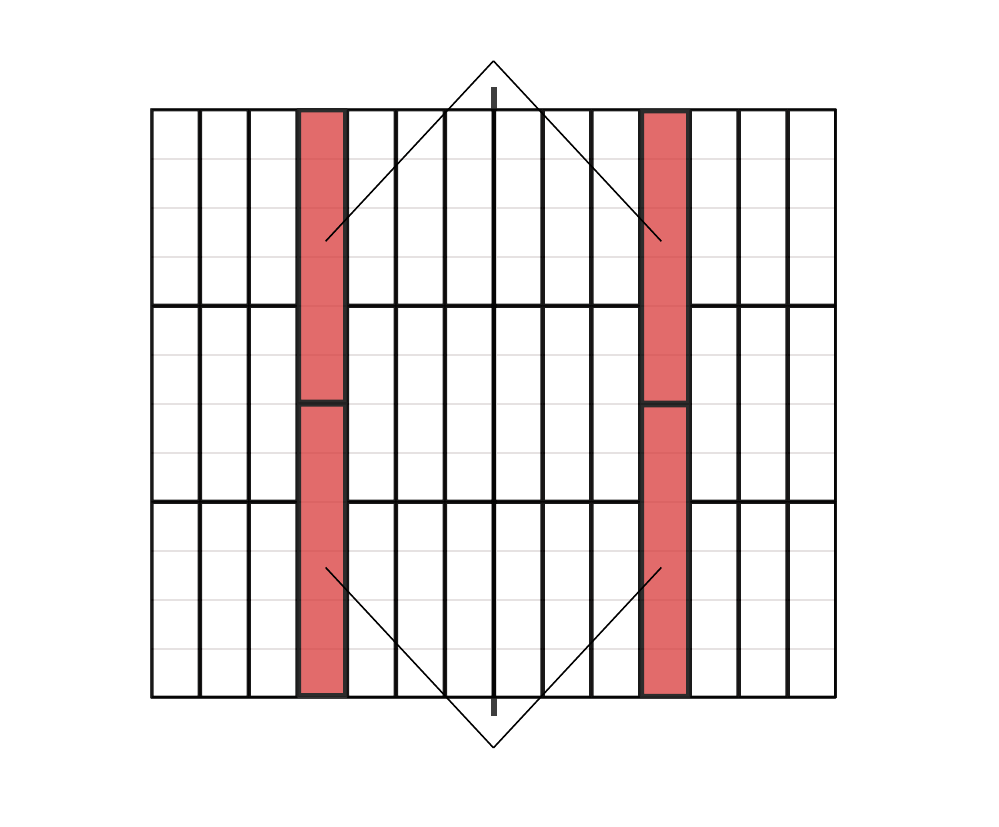}};
		\node at (0, 6.3) [align=center] {\textfonta First pilot-block \par};
		\node at (0, -6.2) [align=center] {\textfonta Second pilot-block \par};
		\end{tikzpicture}
	}
	\caption{Splitting the pilot over two blocks in time for $L_p = 12$.}
	\label{figc4p1:rb02}
%	\vspace{-2mm}
\end{figure}

The easiest solution to the selectivity problem, is to use shorter pilot sequences; however, the shorter the pilot sequences are, the smaller is the number of \acp{UE} that can be supported. Therefore, we have to keep the pilot sequences long enough, in order to support a sufficient number of connections, while at the same time reduce the impact of selectivity. To that end, we consider here repositioning the pilot-blocks around the \acp{RB}. Instead of inserting the pilot sequence along a single \ac{RB} over $12$ consecutive subcarriers, it is split over two neighbouring \acp{RB} in time. This is illustrated in \Cref{figc4p1:rb02}, where a pilot sequence of length $L_p =12$ is split over two blocks, each with a length of six. With such a setup, the pilot length in the frequency direction is halved, and therefore less frequency-selectivity is experienced per sub-sequence. The drawback of such an allocation is the worse time-resolution for the channel estimation, since now only a single effective sequence is used to estimate the channel in time.

\subsection{Example Scenario over TDL-C Channels}
\begin{figure}[t]
	\centering
	\includegraphics[width=1\linewidth]{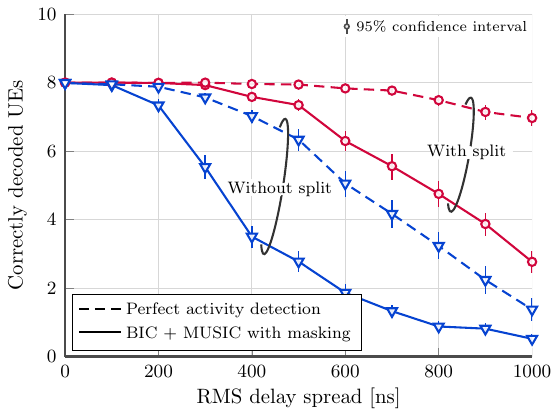}
	\caption{Detection performance over a frequency-selective channel with $K_a = 8$.}
	\label{figc4p1:AD02}
	\vspace{-2mm}
\end{figure}

We investigate the benefit of such an allocation strategy using the same setup of \Cref{Tablec4p1:simpara01}, but we change the channel model to \ac{TDL-C} with varying \ac{RMS} delay spread and fix the average \ac{SNR} of the \acp{UE} to $20$\,dB. The result is shown in \Cref{figc4p1:AD02}.
There are two factors impacting the performance here. On the one hand, due to the increased frequency-selectivity, the performance of the activity detection deteriorates, as we just explained before. On the other hand, the frequency-selectivity impacts the channel estimation performance as well, i.e., as the channel gets more selective, denser pilots are required in order to be able to track the variations in the channel. As can be seen in the figure, splitting the pilot sequence provides gains for both of these issues. The gain for the activity detection can be seen by comparing the performance under \ac{BIC} + \ac{MUSIC}, which shows better robustness against the selectivity with the splitting strategy. The channel estimation gain can be clearly seen when comparing the performance under perfect activity detection. In this case, the split strategy offers better frequency resolution, which helps to provide a better estimation of the channel. Therefore, for such grant-free \ac{NOMA} systems, having the pilot sequences split over multiple \ac{OFDM} symbols might be a good design approach, in order to avoid the aforementioned problems with frequency-selectivity. Of course, this comes at the cost of decreased robustness to time-selectivity. If time-selectivity is an issue, then one solution is to allocate more \ac{OFDM} symbols to the pilots. For example, the 5th and 12th \ac{OFDM} symbols would hold pilot-blocks as well, and then the split can occur over neighbouring \ac{OFDM} symbols only, leading to less time-variation within the split-block. The disadvantage of doing so is a lower data transmission rate.

Finally, it is important to mention that the setup we considered here is a worst-case scenario in which the channels of all the \acp{UE} are suffering from high selectivity. This might not be the case in practice. Also, per 3GPP \cite{3GPP_TR_38900}, $300\,$ns is already considered a long delay spread, and they declare the nominal delay spread to be $100\,$ns. Thus, above $400\,$ns would represent extreme cases. Under moderate conditions, the deterioration in performance may not be severe then, especially when combined with those splitting strategies.

\section{Conclusion}
We investigate the impact of channel correlation on activity detection in grant-free \ac{NOMA} with subspace methods. A setup with a practical frame-structure and where the data and pilot spreading signatures are of different lengths is considered. In the first part, we show how the high correlation of the channel can heavily impact the activity detection performance at the \ac{BS}. In order to mitigate its effect, we propose to overlay the pilot signatures with user-specific masking sequences. This results effectively in a decorrelation of the channel and allows for a successful detection. Then, in the second part, we consider the other extreme where the channel is highly selective, and how it can negatively affect the performance. We show that the impact of high selectivity may be mitigated by a proper reallocation of the pilot signatures.

\begin{acronym}[DSTTDSGRC]
	\setlength{\itemsep}{-3pt}
	\acro{1G}{1st generation}
	\acro{2G}{2nd generation}
	\acro{3G}{3rd generation}
	\acro{3GPP}{Third Generation Partners Project}
	\acro{4G}{4th generation}
	\acro{5G}{5th generation}
%	\acro{5G-NR}{5th generation new-radio}
%	\acro{6G}{sixth-generation}
%	\acro{AD}{activity detection}
	\acro{AIC}{Akaike information criterion}
	\acro{AMP}{approximate message passing}
	\acro{AP}{alternating projection}
	\acro{B5G}{beyond fifth-generation}
%	\acro{BCASC}{best complex antipodal spherical codes}
	\acro{BIC}{Bayesian information criterion}
	\acro{BLER}{block error ratio}
	\acro{BS}{base station}
%	\acro{C-V2X}{cellular-\ac{V2X}}
	\acro{CBGC}{coherence-based Grassmannian codebook}
	\acro{CDF}{cumulative distribution function}
	\acro{CDMA}{code-division multiple-access}
%	\acro{CE}{channel estimation}
	\acro{CLT}{central limit theorem}
	\acro{CRC}{cyclic-redundancy-check}
	\acro{CS}{compressed sensing}
	\acro{CSI}{channel state information}
%	\acro{CSMA}{carrier sense multiple access}
%	\acro{CWL}{codeword level}
%	\acro{DFT}{discrete Fourier transform}
%	\acro{DS}{delay spread}
%	\acro{DSRC}{dedicated short-range communication}
%	\acro{ECDF}{empirical cumulative distribution function}
	\acro{EM}{expectation maximization}
	\acro{EP}{expectation propagation}
	\acro{ETF}{equiangular tight frame}
	\acro{FDMA}{frequency-division multiple access}
	\acro{FFT}{fast-Fourier-transform}
	\acro{IC}{interference cancellation}
	\acro{ICBP}{iterative collision-based packing}
%	\acro{IDMA}{interleave-division multiple access}
%	\acro{IEEE}{institute of electrical and electronics engineers}
%	\acro{IFFT}{inverse fast-Fourier-transform}
%	\acro{IGMA}{interleave-grid multiple access}
	\acro{IoT}{Internet-of-things}
	\acro{IRS}{intelligent reflecting surface}
%	\acro{ITS}{intelligent transport systems}
	\acro{KKT}{Karush-Kuhn-Tucker}
	\acro{LDPC}{low-density parity-check}
%	\acro{LLR}{log-likelhood-ratio}
	\acro{LOS}{line-of-sight}
	\acro{LS}{least-squares}
	\acro{LTE}{Long-Term Evolution}
	\acro{MA}{multiple access}
%	\acro{MAC}{medium access control}
	\acro{MF}{matched filter}
	\acro{MIMO}{multiple-input multiple-output}
	\acro{ML}{maximum likelihood}
	\acro{MMSE}{minimum mean square error}
%	\acro{mMTC}{massive machine-type communication}
	\acro{MRC}{maximum-ratio combining}
	\acro{MTC}{machine-type communication}
	\acro{MUD}{multiuser detection}
%	\acro{MUSA}{multi-user shared access}
	\acro{MUSIC}{MUltiple SIgnal Classification}
	\acro{NLOS}{non-line-of-sight}
	\acro{NOMA}{non-orthogonal multiple access}
	\acro{OFDM}{orthogonal frequency-division multiplexing}
	\acro{OFDMA}{orthogonal frequency-division multiple access}
%	\acro{OGF}{orthoplectic Grassmannian frame}
	\acro{OMA}{orthogonal multiple access}
	\acro{OMP}{orthogonal matching pursuit}
	\acro{PAPR}{peak-to-average-power ratio}
%	\acro{PHY}{physical}
	\acro{PIC}{parallel interference cancellation}
	\acro{QAM}{quadrature amplitude modulation}
	\acro{RB}{resource-block}
	\acro{RE}{resource-element}
	\acro{RIS}{reconfigurable intelligent surface}
	\acro{RMS}{root-mean-square}
	\acro{RV}{random variable}
%	\acro{SB-SPS}{sensing-based semi-persistent scheduling}
	\acro{SDP}{semidefinite programming}
	\acro{SIC}{successive interference cancellation}
	\acro{SINR}{signal-to-interference-plus-noise ratio}
	\acro{SNR}{signal-to-noise ratio}
	\acro{SCMA}{sparse-code multiple access}
	\acro{SVD}{singular value decomposition}
	\acro{TDL-C}{tapped-delay-line-C}
	\acro{TDMA}{time-division multiple access}
	\acro{UE}{user equipment}
%	\acro{V2I}{vehicle-to-infrastructure}
%	\acro{V2N}{vehicle-to-network}
%	\acro{V2P}{vehicle-to-pedestrian}
%	\acro{V2X}{vehicle-to-everything}
%	\acro{V2V}{vehicle-to-vehicle}
	\acro{WBE}{Welch-bound-equality}
\end{acronym}

\bibliographystyle{IEEEtran}
\bibliography{IEEEabrv,./newRefs}

\end{document}